\documentstyle[11pt,newpasp,twoside,epsf]{article}

\begin{document}
\title{Properties of three gas-rich dwarfs in the Centaurus A group}
\author{M. Grossi, M.J. Disney, R.F. Minchin}
\affil{Cardiff University, PO Box 913, Cardiff , UK}
\author{B.J. Pritzl, P.M. Knezek (1), A. Saha}
\affil{National Optical Astronomy Observatories, PO Box 26732, Tucson, AZ 85726, (1) WYIN, Inc, PO Box 26732, Tucson, AZ 85726}
\author{J.S. Gallagher}
\affil{Dept. of Astronomy, University of Wisconsin, Madison, WI
53706-1582}

\author{K.C. Freeman}
\affil{Research School of Astronomy and Astrophysics, Mount Stromlo Observatory, Cotter Road, Weston, ACT 2611, Australia}

\begin{abstract}
We present HST/WFPC2 observations (F555W, F814W) and ATCA high resolution
H{\sc i} maps of three gas-rich dwarf galaxies in the Centaurus A group
 discovered in two blind 21-cm surveys (HIPASS, HIDEEP).
We compare their individual properties and discuss  their star formation
history. Although we can not constrain the age of the oldest
population from the diagrams very well, 
the presence of an extended population
of red giant stars suggests that these systems were
not formed recently. The presence of asymptotic giant branch (AGB) stars in two out
 of three dwarfs, sets a lower limit on the age of about 6 Gyr.
\end{abstract}

\section{Introduction}

The dwarf galaxies in the Local Group (LG) present a wide range of star
 formation histories, gas fractions and metallicities and 
are usually divided in two main classes.
On the one hand gas-deficient, low mass and luminosity  dwarf
spheroidal (dSph) galaxies, generally found within 300 kpc of the
more massive group members. On the other,
 gas-rich dwarf irregular (dIrr) galaxies  have
a larger range in both mass and luminosity and a less clustered distribution.
The richness in gas of a galaxy is defined by the
ratio between its H{\sc i} mass and  B luminosity, expressed in solar units
 ($M_{HI} / L_B$).
Late-type  dIrr in the LG have $M_{HI} / L_B \la 1$. However
galaxies with $M_{HI} / L_B > 1$  can be found in nearby
groups, indicating that a large amount of their gas content has
not yet been processed into stars (C\^ot\'e et al. 1997;  Banks et
al. 1999;  Karachentsev, Karachentseva, \& Huchtmeier 2001). 
Are they recently formed
objects? Or rather systems where star formation (SF) is not
efficient? If so, what inhibits SF in these environments? To
investigate these issues we have selected three gas-rich dwarfs
found in two blind 21-cm surveys, the H{\sc i} Parkes All-Sky
Survey (HIPASS)in the Centaurus A group (Banks et al. 1999) and HIDEEP 
(Minchin et al. 2003). They have faint luminosities ($M_B
\sim -11$) and low surface brightness ($\mu_B > 24$ mag
arcsec$^{-2}$) typical of LG dSphs but with $M_{HI}/L_B$ higher
than  the average value for LG dIrrs (see Table 1). 
Thus, despite their substantial amount 
of gas, that would make them favoured hosts for starbursts, 
their star formation rates (SFRs) tend to be low
($\la 10^{-3} M_{\odot} yr^{-1}$). Being in one of our
closest groups, their stellar population can be resolved with the
use of the Hubble Space Telescope. We present such WFPC2 follow-ups and
 high resolution 21-cm maps taken with the Australian Telescope Compact Array
 (ATCA).

\begin{table}[t]
\caption{Observed properties of the Centaurus A dwarfs}
\begin{tabular}{lcccccc}
\hline 
Object  & $v_{\odot}$ & $M_B$ &  $M_{HI}/M_{\odot}$ & $M_{HI}/L_B $ &  d & D$_{M83}$ \\
 & km s$^{-1}$ & &  & & Mpc & Mpc \\ \hline 
HIPASS J1337-39  &  490  & -11.9 & 3.7$\times 10^7$  &   3.1 & 4.9 & 0.9  \\
HIDEEP J1337-33 & 590 & -10.7 & 5.0$\times 10^6$  &  1.4 & 4.5 & 0.3 \\
HIPASS J1321-31  & 570   & -11.5 & 3.7$\times 10^7$ & 5.1&5.2 & 0.8 \\
\hline 
\end{tabular}
\end{table}

\section{Observations}

The objects
were followed-up with ATCA  in 
two different runs in 2001, using the 750-D and 1.5 km array configurations.
 The total integration time per
 source in each configuration was 12 hours.
In June 2001 the three dwarfs were
observed with the WFPC2 in two
filters (F555W and F814W) for 5000 s and 5200 s respectively.
Details about the observations and data
reduction can be found elsewhere (Pritzl et al. 2003, Grossi et
al. 2003 in prep.).

\section{Results and discussion}

\subsection{Constraining the age of the old stellar population }

The dwarfs show a well developed red giant
branch (RGB) which indicates that the dominant stellar population
consists of intermediate-old age stars. Stars with a wide range of
ages overlay in this area of the diagram making it very difficult to
derive the star formation history from only this feature.
Thus, to constrain the ages of these galaxies we can only  use
the AGB stars, indicators of the intermediate-to-old
 age
stellar population (less than 10 Gyr). There is evidence of
stars that lie above the tip of the RGB (at 1.0 $< V -
I < $ 1.3) in {\bf HIPASS J1337-39} (Fig. 1). The lack of stars in the field
with that range of colors and magnitudes suggests that these are AGB stars
in J1337-39.
Theoretical isochrones from the Padua group (Girardi et al. 2002,
Bertelli et al. 1994) can match the position of these stars
assuming an age of $\sim $ 6 Gyr and a metallicity of Z=
0.0004 (1/50 solar). A few  AGB candidates can also be identified
in {\bf HIDEEP J1337-33} although less numerous than in J1337-39.
Again comparison with stellar tracks gives an age of $\sim 6$ Gyr
for a metallicity Z = 0.001 (1/20 solar). {\bf HIPASS J1321-31}
does not seem to show evident AGB stars. However its color magnitude
diagram (CMD) presents
a peculiar bright `red plume' extending up to I $\sim$ 22.6. We
have ruled out the possibility of this being the RGB of the galaxy
(Pritzl et al. 2003) and we believe that it is more likely related
to recent ($\la $ 1 Gyr) SF activity in this galaxy. We briefly
discuss our interpretation in the next section.

\begin{figure}[t]
\plotfiddle{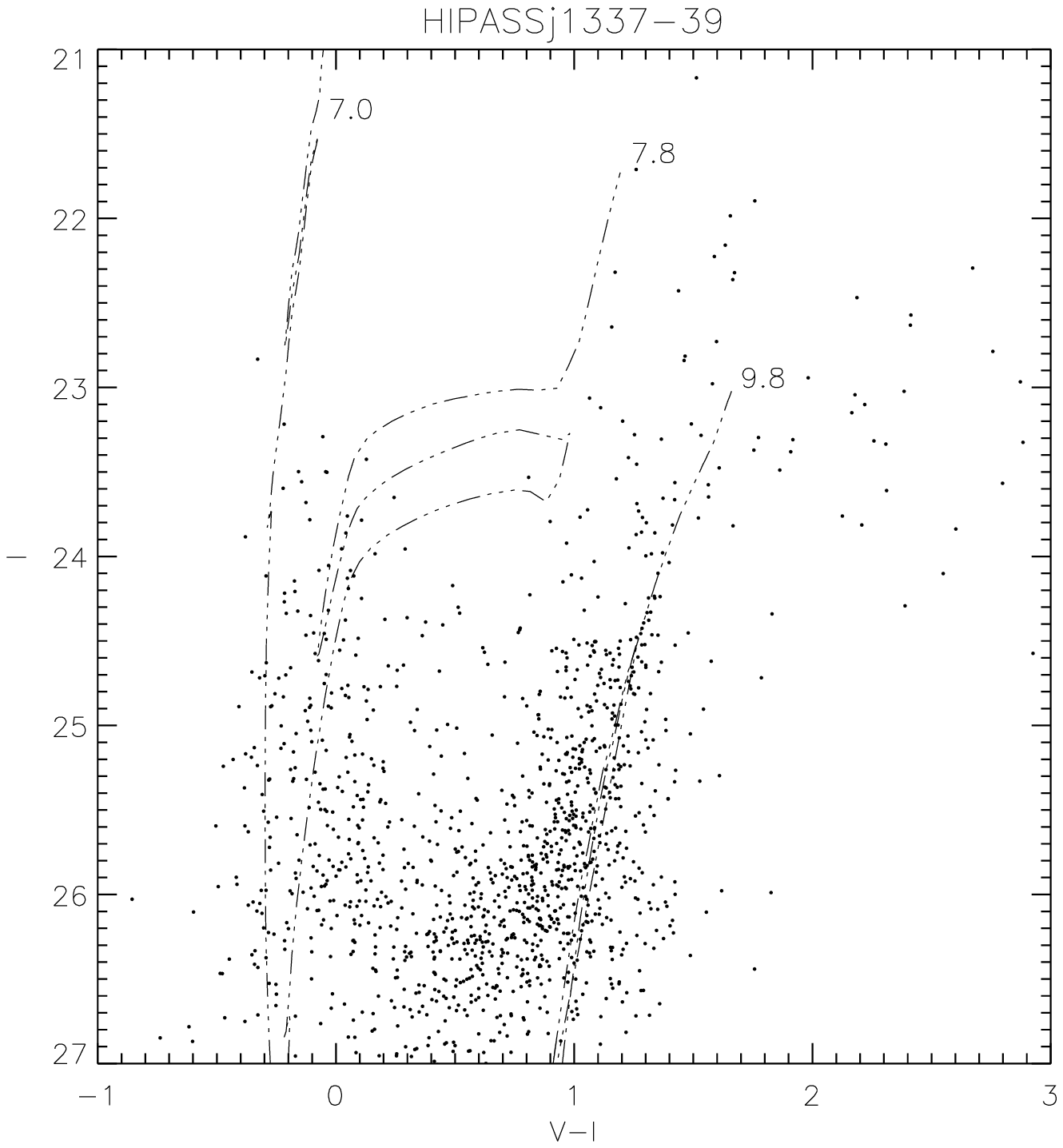}{2.0cm}{0.}{34.}{32.}{-265}{-130}
\plotfiddle{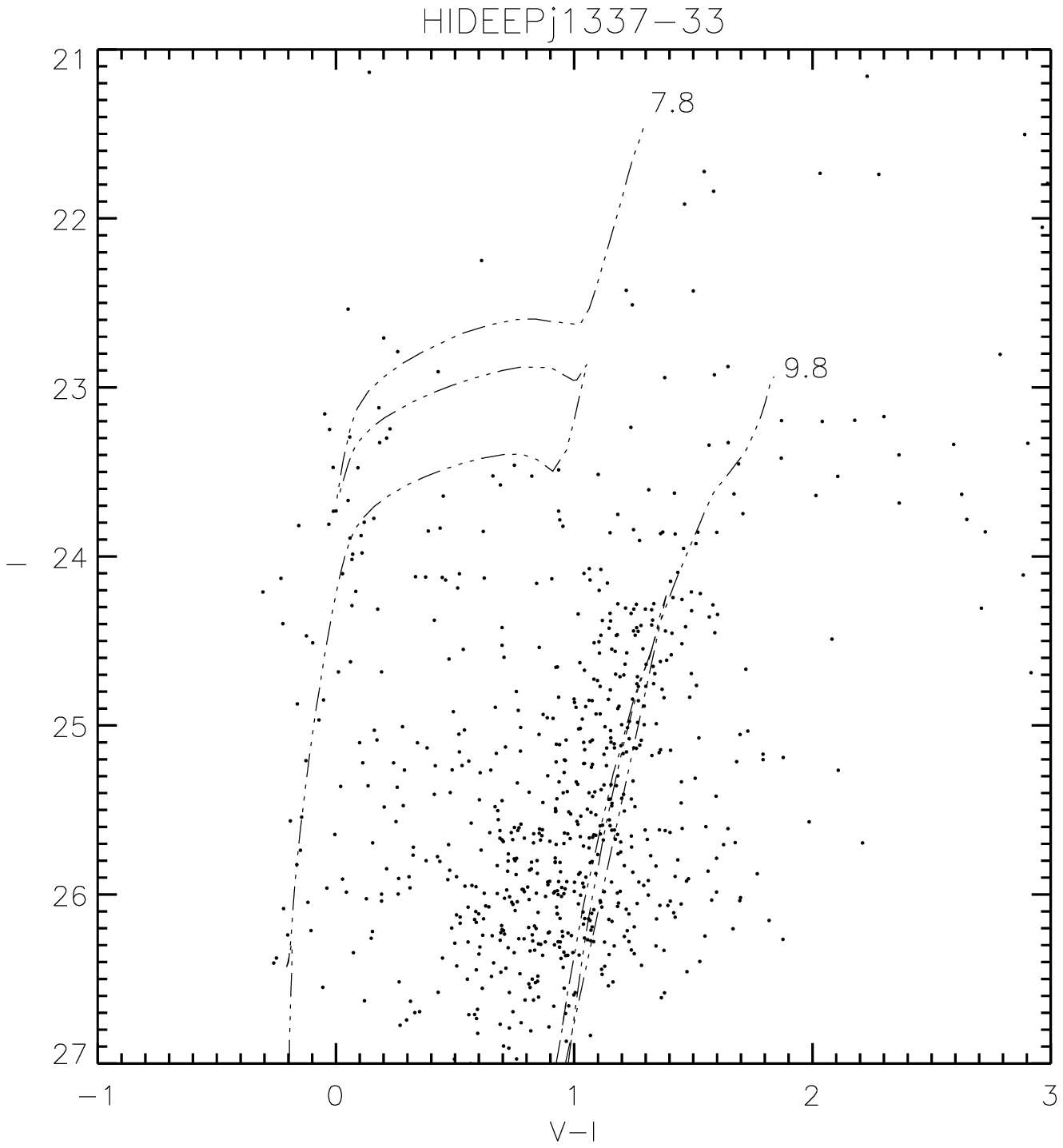}{0cm}{0.}{34.}{32.}{-130}{-105}
\plotfiddle{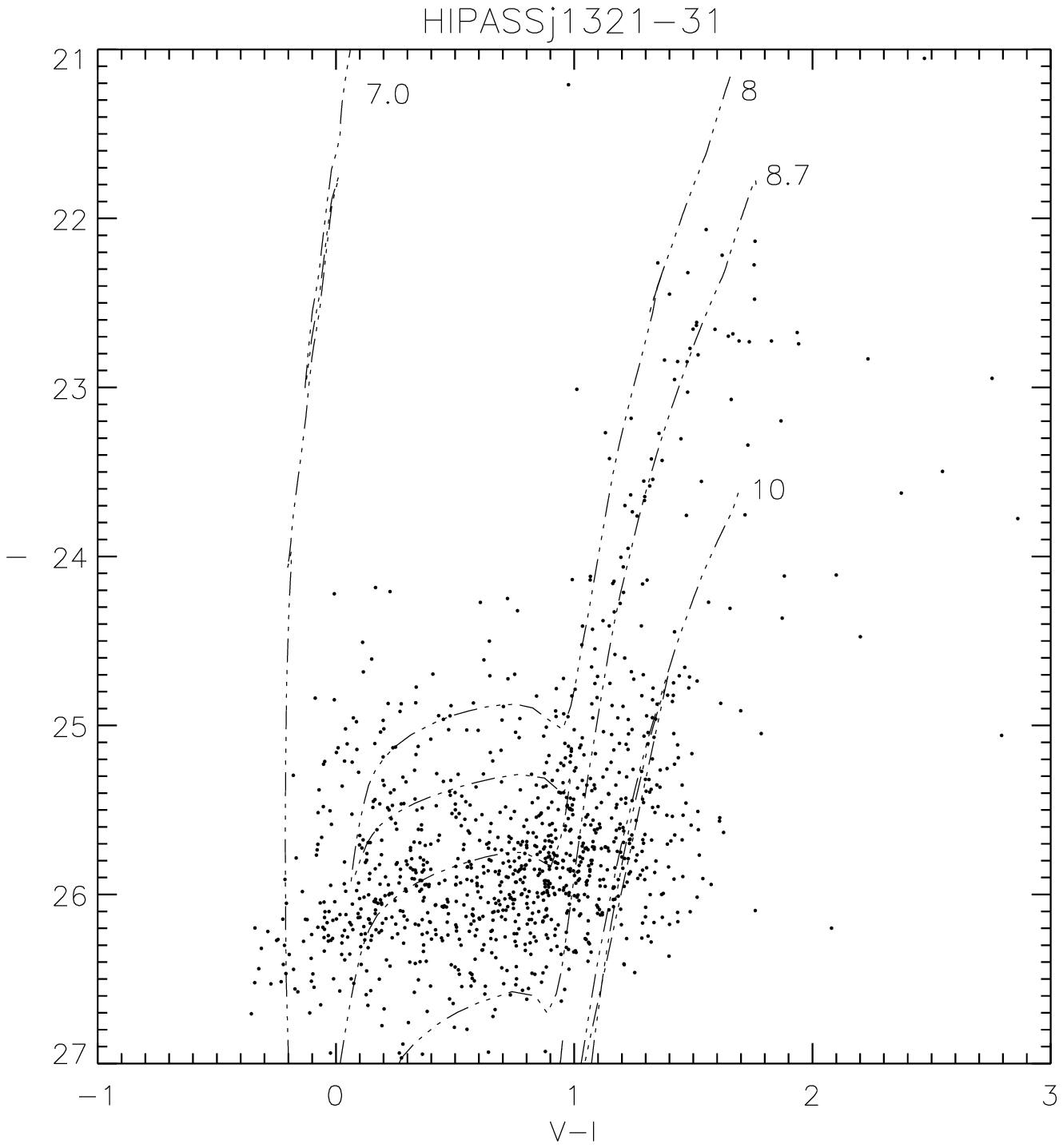}{0cm}{0.}{34.}{32.}{10}{-80}
\plotfiddle{figure4.eps}{2.7cm}{-90.}{24}{24}{-252.5}{80}
\plotfiddle{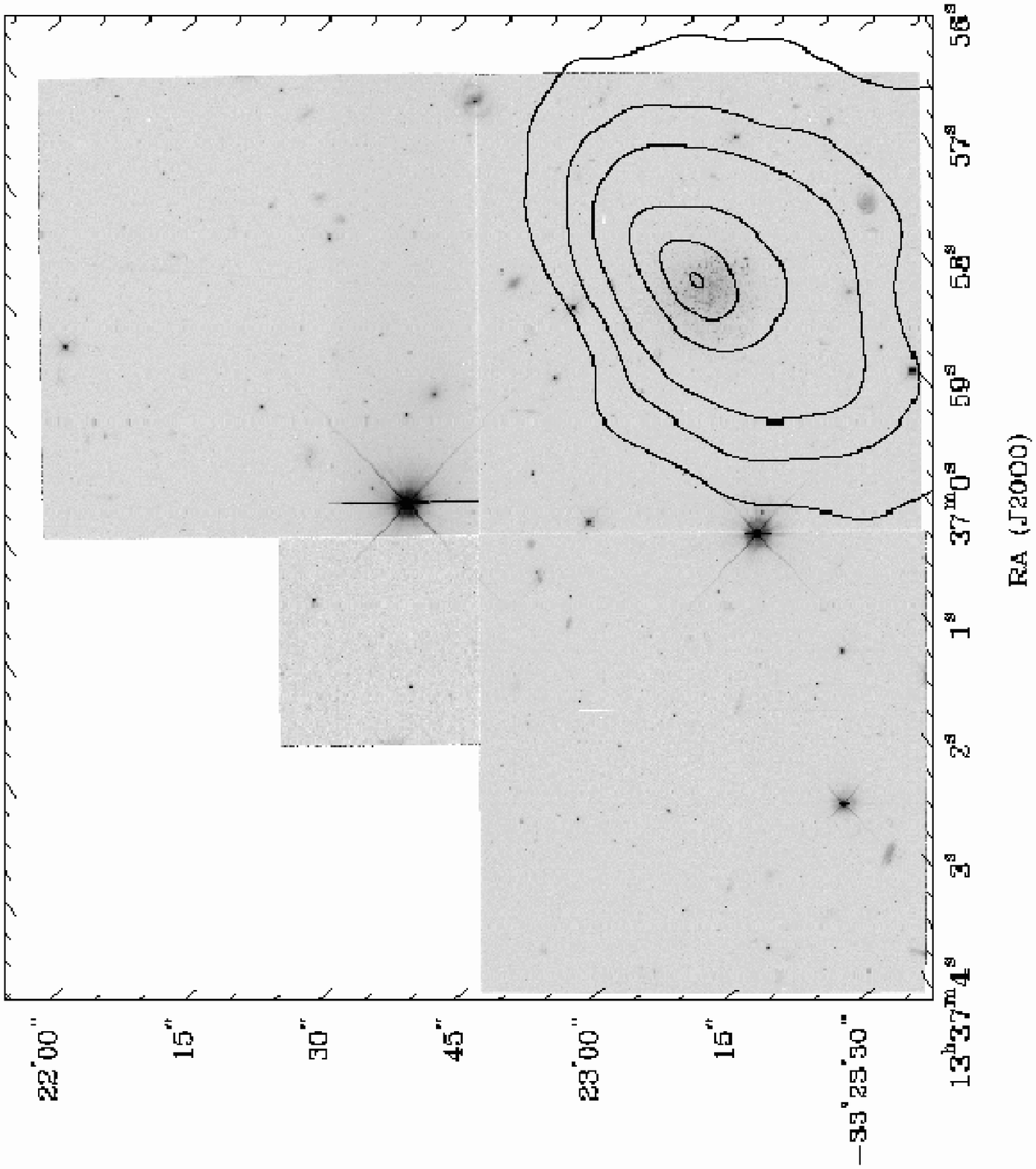}{0.cm}{-90.0}{21.35}{21.4}{-110}{98}
\plotfiddle{figure6.eps}{0.cm}{-90.0}{24.}{24.}{22.5}{130}
\caption{{\bf left)} HIPASS J1337-39: {\bf up)} (V-I), I
color magnitude diagram with Padua group theoretical isochrones
overlaid at z=0.0004 (1/50 solar). The numbers indicate the
corresponding log $Age$. {\bf down)} H{\sc i} column density
contours overlaid on the DSS image of the field. Levels: 2.5, 5.0
7.5, 10, 12.5, 15.0, 17.5 $\times 10^{20}$ cm$^{-2}$. {\bf
center)} HIDEEP J1337-33: {\bf up)} (V-I), I color magnitude
diagram with Padua isochrones at z=0.001 (1/20 solar). {\bf down)}
H{\sc i} column density contours overlaid on the WFPC2 field.
Levels: 0.5, 1.0 1.5, 2.0, 2.25, 2.35 $\times 10^{20}$ cm$^{-2}$.
{\bf right)} HIPASS J1321-31: {\bf up)} (V-I), I color
magnitude diagram with Padua stellar tracks at z=0.0004. {\bf
down)} H{\sc i} column density contours overlaid on the DSS field.
Levels: 0.5, 1.0 1.5, 2.0, 2.5 $\times 10^{20}$ cm$^{-2}$. }
\end{figure}

\subsection{The more recent SF activity and the gas distribution}

{\bf HIPASS J1337-39} is the only galaxy which is currently forming
stars as one H{\sc ii} regions can be found in the southern part of the
galaxy. The region correlates with the peak in the H{\sc i}
column density ($N_{H{\sc i}} = 2 \times 10^{21}$ cm$^{-2}$), the highest
among the three dwarfs. The well developed
blue plume at  ($V - I$) $< 0.2$,
indicates a population of core-helium burning stars with an age of
60-100 Myr (Fig. 1).
The small number of blue stars (at -0.2 $< V - I <$ 0.5) in {\bf
HIDEEP J1337-33} and the absence of evident H{\sc ii} regions  indicates
a drop in the star formation activity  at
around 60 Myr (Fig. 1). The neutral gas density in the optical
extension of the galaxy is almost constant, hovering around
$N_{H{\sc i}} = 2 \times 10^{20}$ cm$^{-2}$.
There is no evidence of very recent SF in {\bf HIPASS J1321-31} either.
However the thin red
plume in its CMD indicates a peculiar star formation history (SFH).
We have proposed that it consists of  core-helium burning stars
in the red super giant (RSG) phase (Pritzl et al. 2003).
This scenario would imply that
the galaxy went through a period of enhanced star formation
activity less than 1 Gyr ago. The more massive stars would have 
already vacated the blue spike while the RSG branch  would presently
be inhabited by $\la 3 M_{\odot}$ stars, possibly 500 Myr old assuming a
metallicity of 1/50 solar (Fig. 1). To explain the
presence of the faint blue stars, the SF activity must have continued
after the burst at
a decreasing rate, and dropped off around 100 Myr ago. The H{\sc i}
distribution (Fig. 1) is offset from the optical center with an overall low
 gas column density whose peaks (at $N_{H{\sc i}} = 2.5 \times 10^{20}$
cm$^{-2}$) do not seem to be related to the main optical
counterpart.

\section{Conclusions}

We can rule out the possibility that the HIPASS dwarfs in
Cen A are recently formed objects as indicated by the presence of
a population of  RGB and AGB stars.
However the issue of their low star formation rates is still open.
The overall low gas column density, apart from sporadic local
enhancements, may be one possible explanation, although  we can
not exclude connections with the local environment. M83 is the
closest massive galaxies to the dwarfs, but only HIDEEP
J1337-33 is within 300 kpc from it (see Table 1). On the one hand
the lack of interactions may prevent massive SF from occuring in the
more isolated dwarfs like
 J1337-39 or J1321-31. On the other, gas-stripping may be responsible
for the current quiescence of J1337-33, a possible satellite of M83.

\end{document}